\documentclass[%
 reprint,
 xcolor,
 amsmath,amssymb,esint,commath,
 aps,
 prl,
]{revtex4-1}

\usepackage{graphicx}
\usepackage{dcolumn}
\usepackage{bm}
\usepackage{hyperref}
\usepackage[mathlines]{lineno}
\usepackage{float}

\newcommand*\dif{\mathop{}\!\mathrm{d}} %

\newcommand{\FDModeFreq}{15,538\,Hz}

\newcommand{\FDModeGroupAFreq}{15.00\,kHz}

\newcommand{\FDModeGroupEFreq}{15.54\,kHz}

\newcommand{\FDRingupTau}{182\,sec}
\newcommand{\FDRingdownTau}{23\,sec}
\newcommand{\FDRingupTauwU}{$182\pm 9$\,sec}
\newcommand{\FDRingdownTauwU}{$23\pm 1$\,sec}
\newcommand{\FDR}{2.4}
\newcommand{\FDReff}{0.18}
\newcommand{\FDRwU}{$2.4\pm 0.8$}
\newcommand{\FDReffwU}{$0.18\pm 0.06$}

\newcommand{\FDStartDampingForce}{0.62\,nN rms}
\newcommand{\FDDampedStateAverageRMS}{0.03\,nN rms}
\newcommand{\FDDampedStatePeak}{0.11\,nN peak} 

\newcommand{\FDHighPowerDSAverageRMS}{1.5\,nN rms} 

\newcommand{\FDAverageRMSDSVoltage}{0.42\,mV rms} 
\newcommand{\FDPeakDSVoltage}{1.4\,mV peak} 

\newcommand{\FDSafetyFactor}{10,000} 
\newcommand{\FDSafetyFactorHP}{310} 
\newcommand{\FDSafetyFactorOM}{10} 

\begin{document}

\preprint{APS/123-QED}

\title{First Demonstration of Electrostatic Damping of Parametric Instability at Advanced LIGO}
\author{Carl Blair$^1$}
\email{carl.blair@uwa.edu.au}
\author{Slawek Gras$^2$}
\author{Richard Abbott$^5$}
\author{Stuart Aston$^3$}
\author{Joseph Betzwieser$^3$}
\author{David Blair$^1$}
\author{Ryan DeRosa$^3$}
\author{Matthew Evans$^2$}
\author{Valera Frolov$^3$}
\author{Peter Fritschel$^2$}
\author{Hartmut Grote$^4$}
\author{Terra Hardwick$^5$}
\author{Jian Liu$^1$}
\author{Marc Lormand$^3$}
\author{John Miller$^2$}
\author{Adam Mullavey,$^3$ } 
\author{Brian O'Reilly$^3$}
\author{Chunnong Zhao$^1$}
\affiliation{%
$^1$ University of Western Australia, Crawley, Western Australia 6009, Australia
}%
\affiliation{
$^2$ Massachusetts Institute of Technology, Cambridge, Massachusetts 02139, USA
}
\affiliation{
$^3$LIGO Livingston Observatory, Livingston, Louisiana 70754, USA
}%
\affiliation{
$^4$Max Planck Institute for Gravitational Physics, 30167 Hannover, Germany
}%
\affiliation{
$^5$California Institute of Technology, Pasadena 91125, USA \\
}
\affiliation {$^6$Louisiana State University, Baton Rouge, Louisiana 70803, USA }

\author{
B.~P.~Abbott,$^{1}$  
T.~D.~Abbott,$^{2}$  
C.~Adams,$^{3}$  
R.~X.~Adhikari,$^{1}$  
S.~B.~Anderson,$^{1}$  
A.~Ananyeva,$^{1}$  
S.~Appert,$^{1}$  
K.~Arai,$^{1}$	
S.~W.~Ballmer,$^{4}$  
D.~Barker,$^{5}$
% ...
\newline
\medskip
\centerline{(LSC Instrument Authors	)}\noaffiliation
\medskip
\parindent 0pt
\medskip
}\noaffiliation
\affiliation {LIGO, California Institute of Technology, Pasadena, CA 91125, USA }
\affiliation {Louisiana State University, Baton Rouge, LA 70803, USA }
\affiliation {American University, Washington, D.C. 20016, USA }
\affiliation {University of Florida, Gainesville, FL 32611, USA }
\affiliation {LIGO Livingston Observatory, Livingston, LA 70754, USA }
\affiliation {University of Sannio at Benevento, I-82100 Benevento, Italy and INFN, Sezione di Napoli, I-80100 Napoli, Italy }
\affiliation {Albert-Einstein-Institut, Max-Planck-Institut f\"ur Gravi\-ta\-tions\-physik, D-30167 Hannover, Germany }
\affiliation {LIGO, Massachusetts Institute of Technology, Cambridge, MA 02139, USA }
\affiliation {Instituto Nacional de Pesquisas Espaciais, 12227-010 S\~{a}o Jos\'{e} dos Campos, S\~{a}o Paulo, Brazil }
\affiliation {Inter-University Centre for Astronomy and Astrophysics, Pune 411007, India }
\affiliation {International Centre for Theoretical Sciences, Tata Institute of Fundamental Research, Bangalore 560012, India }
\affiliation {University of Wisconsin-Milwaukee, Milwaukee, WI 53201, USA }
\affiliation {Leibniz Universit\"at Hannover, D-30167 Hannover, Germany }
\affiliation {Australian National University, Canberra, Australian Capital Territory 0200, Australia }
\affiliation {The University of Mississippi, University, MS 38677, USA }
\affiliation {California State University Fullerton, Fullerton, CA 92831, USA }
\affiliation {Chennai Mathematical Institute, Chennai 603103, India }
\affiliation {University of Southampton, Southampton SO17 1BJ, United Kingdom }
\affiliation {Universit\"at Hamburg, D-22761 Hamburg, Germany }
\affiliation {Albert-Einstein-Institut, Max-Planck-Institut f\"ur Gravitations\-physik, D-14476 Potsdam-Golm, Germany }
\affiliation {Montana State University, Bozeman, MT 59717, USA }
\affiliation {Syracuse University, Syracuse, NY 13244, USA }
\affiliation {SUPA, University of Glasgow, Glasgow G12 8QQ, United Kingdom }
\affiliation {LIGO Hanford Observatory, Richland, WA 99352, USA }
\affiliation {Columbia University, New York, NY 10027, USA }
\affiliation {Stanford University, Stanford, CA 94305, USA }
\affiliation {Center for Relativistic Astrophysics and School of Physics, Georgia Institute of Technology, Atlanta, GA 30332, USA }
\affiliation {University of Birmingham, Birmingham B15 2TT, United Kingdom }
\affiliation {RRCAT, Indore MP 452013, India }
\affiliation {Faculty of Physics, Lomonosov Moscow State University, Moscow 119991, Russia }
\affiliation {SUPA, University of the West of Scotland, Paisley PA1 2BE, United Kingdom }
\affiliation {University of Western Australia, Crawley, Western Australia 6009, Australia }
 \affiliation {Washington State University, Pullman, WA 99164, USA }
 \affiliation {Embry-Riddle Aeronautical University, Prescott, AZ 86301, USA }
\affiliation {University of Oregon, Eugene, OR 97403, USA }
\affiliation {Carleton College, Northfield, MN 55057, USA }
\affiliation {University of Maryland, College Park, MD 20742, USA }
\affiliation {NASA/Goddard Space Flight Center, Greenbelt, MD 20771, USA }
\affiliation {Canadian Institute for Theoretical Astrophysics, University of Toronto, Toronto, Ontario M5S 3H8, Canada }
\affiliation {Tsinghua University, Beijing 100084, China }
\affiliation {Texas Tech University, Lubbock, TX 79409, USA }
\affiliation {National Tsing Hua University, Hsinchu City, 30013 Taiwan, Republic of China }
\affiliation {Charles Sturt University, Wagga Wagga, New South Wales 2678, Australia }
\affiliation {West Virginia University, Morgantown, WV 26506, USA }
\affiliation {University of Chicago, Chicago, IL 60637, USA }
\affiliation {Caltech CaRT, Pasadena, CA 91125, USA }
\affiliation {Korea Institute of Science and Technology Information, Daejeon 305-806, Korea }
\affiliation {University of Brussels, Brussels 1050, Belgium }
\affiliation {Sonoma State University, Rohnert Park, CA 94928, USA }
\affiliation {Center for Interdisciplinary Exploration \& Research in Astrophysics (CIERA), Northwestern University, Evanston, IL 60208, USA }
\affiliation {University of Minnesota, Minneapolis, MN 55455, USA }
\affiliation {The University of Melbourne, Parkville, Victoria 3010, Australia }
\affiliation {Institute for Plasma Research, Bhat, Gandhinagar 382428, India }
\affiliation {The University of Sheffield, Sheffield S10 2TN, United Kingdom }
\affiliation {The University of Texas Rio Grande Valley, Brownsville, TX 78520, USA }
\affiliation {The Pennsylvania State University, University Park, PA 16802, USA }
\affiliation {Cardiff University, Cardiff CF24 3AA, United Kingdom }
\affiliation {Montclair State University, Montclair, NJ 07043, USA }
\affiliation {MTA E\"otv\"os University, ``Lendulet'' Astrophysics Research Group, Budapest 1117, Hungary }
\affiliation {School of Mathematics, University of Edinburgh, Edinburgh EH9 3FD, United Kingdom }
\affiliation {Indian Institute of Technology, Gandhinagar Ahmedabad Gujarat 382424, India }
\affiliation {University of Szeged, D\'om t\'er 9, Szeged 6720, Hungary }
\affiliation {Tata Institute of Fundamental Research, Mumbai 400005, India }
\affiliation {University of Michigan, Ann Arbor, MI 48109, USA }
\affiliation {Rochester Institute of Technology, Rochester, NY 14623, USA }
\affiliation {University of Massachusetts-Amherst, Amherst, MA 01003, USA }
\affiliation {Universitat de les Illes Balears, IAC3---IEEC, E-07122 Palma de Mallorca, Spain }
\affiliation {SUPA, University of Strathclyde, Glasgow G1 1XQ, United Kingdom }
\affiliation {IISER-TVM, CET Campus, Trivandrum Kerala 695016, India }
\affiliation {Institute of Applied Physics, Nizhny Novgorod, 603950, Russia }
\affiliation {Pusan National University, Busan 609-735, Korea }
\affiliation {Hanyang University, Seoul 133-791, Korea }
\affiliation {University of Adelaide, Adelaide, South Australia 5005, Australia }
\affiliation {Monash University, Victoria 3800, Australia }
\affiliation {Seoul National University, Seoul 151-742, Korea }
\affiliation {University of Alabama in Huntsville, Huntsville, AL 35899, USA }
\affiliation {Southern University and A\&M College, Baton Rouge, LA 70813, USA }
\affiliation {College of William and Mary, Williamsburg, VA 23187, USA }
\affiliation {Instituto de F\'\i sica Te\'orica, University Estadual Paulista/ICTP South American Institute for Fundamental Research, S\~ao Paulo SP 01140-070, Brazil }
\affiliation {University of Cambridge, Cambridge CB2 1TN, United Kingdom }
\affiliation {IISER-Kolkata, Mohanpur, West Bengal 741252, India }
\affiliation {Rutherford Appleton Laboratory, HSIC, Chilton, Didcot, Oxon OX11 0QX, United Kingdom }
\affiliation {Whitman College, 345 Boyer Avenue, Walla Walla, WA 99362 USA }
\affiliation {National Institute for Mathematical Sciences, Daejeon 305-390, Korea }
\affiliation {Hobart and William Smith Colleges, Geneva, NY 14456, USA }
\affiliation {King's College London, University of London, London WC2R 2LS, United Kingdom }
\affiliation {Andrews University, Berrien Springs, MI 49104, USA }
\affiliation {Trinity University, San Antonio, TX 78212, USA }
\affiliation {University of Washington, Seattle, WA 98195, USA }
\affiliation {Kenyon College, Gambier, OH 43022, USA }
\affiliation {Abilene Christian University, Abilene, TX 79699, USA }

\collaboration{LSC Collaboration}



\date{\today}

\begin{abstract}
Interferometric gravitational wave detectors operate with high optical power in their arms in order to achieve high shot-noise limited strain sensitivity. A significant limitation to increasing the optical power is the phenomenon of three-mode parametric instabilities, in which the laser field in the arm cavities is scattered into higher order optical modes by acoustic modes of the cavity mirrors. The optical modes can further drive the acoustic modes via radiation pressure, potentially  producing an exponential buildup. One proposed technique to stabilize parametric instability is active damping of acoustic modes. We report here the first demonstration of damping a parametrically unstable mode using active feedback forces on the cavity mirror. A \FDModeFreq\ mode that grew exponentially with a time constant of \FDRingupTau\ was damped using electro-static actuation, with a resulting decay time constant of \FDRingdownTau. An average control force of \FDDampedStateAverageRMS\ was required to maintain the acoustic mode at its minimum amplitude.
\newpage

\end{abstract}

\pacs{Valid PACS appear here}
\maketitle

\textit{Introduction} Three-mode parametric instability (PI) has been a known issue for advanced laser interferometer gravitational wave detectors since first recognised by Braginsky et al \cite{Braginsky}, and modelled in increasing detail \cite{ZhaoPI, Strigin2008, EvansPIgeneral, Gras2010, Vyatchanin12}. The phenomenon was first observed in 2009 in microcavities \cite{Tomes2009}, then in 2014 in an 80\,m cavity \cite{ginginPI} and soon afterwards during the commissioning of Advanced LIGO \cite{EvansPIobs}.  Left uncontrolled PI results in the optical cavity control systems becoming unstable on time scales of tens of minutes to hours \cite{EvansPIobs}.
 
The first detection of gravitational waves was made by two Advanced LIGO laser interferometer gravitational wave detectors with about 100\,kW of circulating power in their arm cavities \cite{GWDectection}.  To achieve this power level required suppression of PI through thermal tuning of the higher-order mode eigen-frequency \cite{Zhao2005} explained later in this paper. 
This tuning allowed the optical power to be increased in Advanced LIGO from about 5\,\% to 12\,\% of the design power, sufficient to attain a strain sensitivity of $10^{-23}\rm \,Hz^{-\frac{1}{2}}$ at 100\,Hz. 

At the design power it will not be possible to avoid instabilities using thermal tuning alone for two reasons.  First the parametric gain scales linearly with optical power and second the acoustic mode density is so high that thermal detuning for one acoustic mode brings other modes into resonance \cite{Zhao2005,EvansPIobs}.
 
Several methods are likely to be useful for controlling PI.  Active thermal tuning will minimize the effects of thermal transients \cite{Fan2008,ramette16} and maintain operation near the parametric gain minimum. In the future, acoustic mode dampers attached to the test masses \cite{grasDampers2016} could damp  acoustic modes.  Active damping \cite{Electrostatic} of acoustic modes can also suppress instabilities, by
applying feedback forces to the test masses. 
 
In this letter we report on the control of a PI by actively damping a \FDModeGroupEFreq\ acoustic mode of an Advanced LIGO test mass using electro-static force actuators.  
First we review the physics of PI and the status of PI control in LIGO. Then we discuss the electrostatic drive system at LIGO and how it interacts with the test mass modes. 
Then we summarise the experimental configuration, report successful damping observations, and discuss the implications for high power operation of Advanced LIGO.
  
\textit{Parametric Instability} The parametric gain $R_m$, as derived by Evans et al \cite{EvansPIgeneral} is given by;
\begin{equation}
R_{\rm m} = \frac{ 8 \pi Q_m P}{M\omega_m^2c\lambda_0}\sum_{n=1}^{\infty} \mathcal{R}e[G_n]B_{m,n}^2 \label{eq:parametricGain}
\end{equation}
Here $Q_{m}$ is the quality factor (Q) of the mechanical mode $m$,  $P$ is the power in the fundamental optical mode of the cavity, $M$ is the mass of the test mass, $c$ is the speed of light, $\lambda_0$ is the wavelength of light, $\omega_{m}$ is the mechanical mode angular frequency, $G_n$ is the transfer function for an optical field leaving the test mass surface to the field incident on that same surface and $B_{m,n}$ is the spatial overlap between the optical beat note pressure distribution and the mechanical mode surface deformation. 

It is instructive to consider the simplified case of a single cavity and a single optical mode to understand the phenomena.  For a simulation analysis including arms and recycling cavities see~\cite{Gras2010, EvansPIgeneral} and for an explanation of dynamic effects that may make high parametric gains from the recycling cavities less likely see~\cite{ginginPI}.  In the simplified case we consider the $TEM_{\rm 03}$ mode as it dominates the optical interaction with the acoustic mode investigated here;
\begin{equation}
 \mathcal{R}e[G_{03}] = \frac{c}{L\pi\gamma(1+\Delta\omega^2/\gamma^2)} \label{eq:OpticalTF}
\end{equation}
Here $\gamma$ is the half-width at half maximum of the $TEM_{\rm 03}$ optical mode frequency distribution, L is the length of the cavity, $\Delta\omega$ is the spacing in frequency between the mechanical mode $\omega_m$ and the beat note of the fundamental and TEM$_{03}$ optical modes.  
In general the parametric gain changes the time constant of the mechanical mode as in Equation~\ref{eq:piEffQ}. If the parametric gain exceeds unity the mode becomes unstable.
\begin{equation}
\tau_{\rm pi} = \tau_{\rm m} / (1 - R_{\rm m}) \label{eq:piEffQ}
\end{equation}

\begin{figure}[hb]
\centering\includegraphics[width=0.99\linewidth]{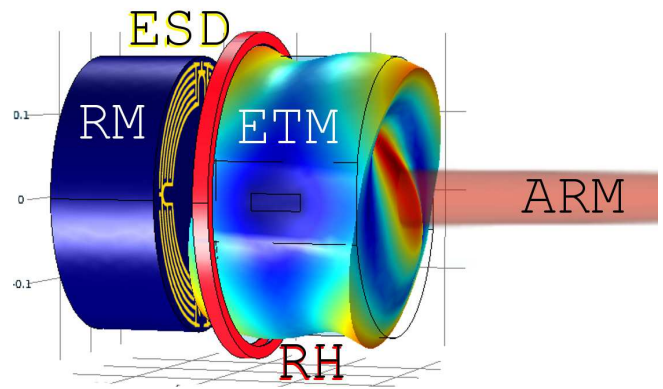}
\caption{Schematic of the gold ESD comb on the  reaction mass (RM), the ring heater (RH) and the end test mass (ETM) with exaggerated deformation due to the \FDModeFreq\ mode.  The colour represents the magnitude of the displacement (red is large, blue is small).  The laser power in the arm cavity is depicted in red (ARM). Suspension structures are not shown and while the scale is marked to the left the  distance between RM and ETM is exaggerated by a factor of 10} \label{fig:ESDlocation}
\end{figure}

Where $\tau_m$ is the natural time constant of the mechanical mode and $\tau_{pi}$ is the time constant of the mode influenced by the opto-mechanical interaction.  Thermal tuning was used to control PI in Advanced LIGO's Observation run 1 and was integral to this experiment, so will be examined in some detail.  
Thermal tuning is achieved using radiative ring heaters that surround the barrel of each test mass without physical contact as in Figure~\ref{fig:ESDlocation}.  Applying power to the ring heater decreases the radius of curvature (RoC) of the mirrors. This changes the cavity g-factor and tunes the mode spacing between the fundamental ($TEM_{00}$) and higher order transverse electromagnetic ($TEM_{mn}$) modes in the cavity, thereby tuning the parametric gain by changing $\Delta\omega$ in Equation~\ref{eq:OpticalTF}. 

\begin{figure*}
\centering \includegraphics[width=0.99\linewidth]{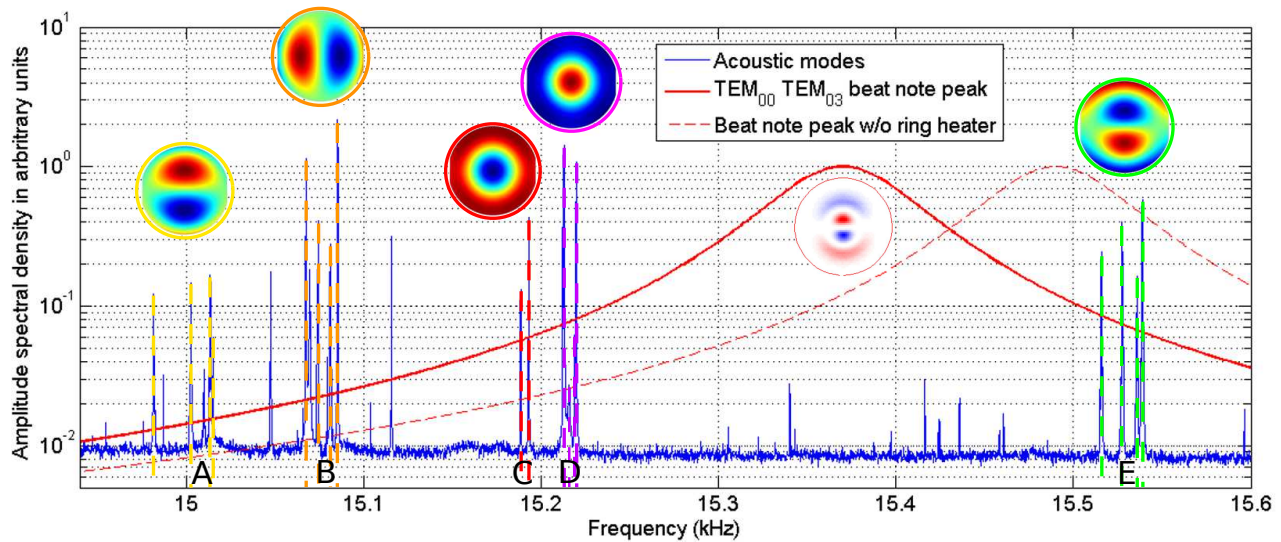}
\caption{The relative location of the optical and mechanical modes during Advanced LIGO Observation run 1.  Mechanical modes measured in transmission of the Output mode cleaner shown in blue with mode surface deformation generated from FEM modeling overlay-ed.  These modes appear in groups of  four, one for each test mass. They have line-width $\sim 1mHz$.  The beat note between the fundamental $TEM_{\rm 00}$ and $TEM_{\rm 03}$ optical cavity modes for a simplified single cavity is shown in bold red and with the ring heater turned off, in dashed red.  The shape of the $TEM_{\rm 03}$ mode simulated with OSCAR \cite{OSCAR} is inset below the peak.}  \label{fig:modeCartoon} 
\end{figure*}

Figure~\ref{fig:modeCartoon} shows the optical gain curve (Equation~\ref{eq:OpticalTF}) for the $\rm TEM_{03}$ mode, with the ring heater tuning used during Advanced LIGO’s first observing run \cite{LigoO1}.  
With no thermal tuning, the optical gain curve in Figure~\ref{fig:modeCartoon} moves to higher frequency, decreasing the frequency spacing $\Delta\omega$ with mode group E. 
This leads to the instability of this group of modes. (Note that the mirror acoustic mode frequencies are only weakly tuned by heater power, due to the small value of the fused silica temperature dependence of  Young's modulus).
If the ring heater power is increased inducing approximately 5\,m change in radius of curvature, the beat note gain curve in Figure~\ref{fig:modeCartoon} moves left about 400\,Hz, decreasing the value $\Delta\omega$ for mode group A, resulting in their instability.  
The mode groups C and D are stable as the second and fourth order optical modes that might be excited from these modes are far from resonance.  Mode Group B is also stable at the circulating optical power used in this experiment presumably due to either lower quality factor $Q_m$ or lower optical gain $G_{30}$ of the TEM$_{30}$ mode as investigated in \cite{Barriga07}.  
If the power in the interferometer is increased by a factor of 3 there will no longer be a stable region.  Mode group A at \FDModeGroupAFreq\ and group E at \FDModeGroupEFreq\ will be unstable simultaneously. 

\textit{Electrostatic Control} Electrostatic control of PI was proposed \cite{Ju2009} and studied in the context of the LIGO electrostatic control combs by Miller et al \cite{Electrostatic}.  Here we report studies of electrostatic feedback damping for the group E modes at \FDModeGroupEFreq.   

\begin{figure}[ht]
\centering\includegraphics[width=0.44\linewidth]{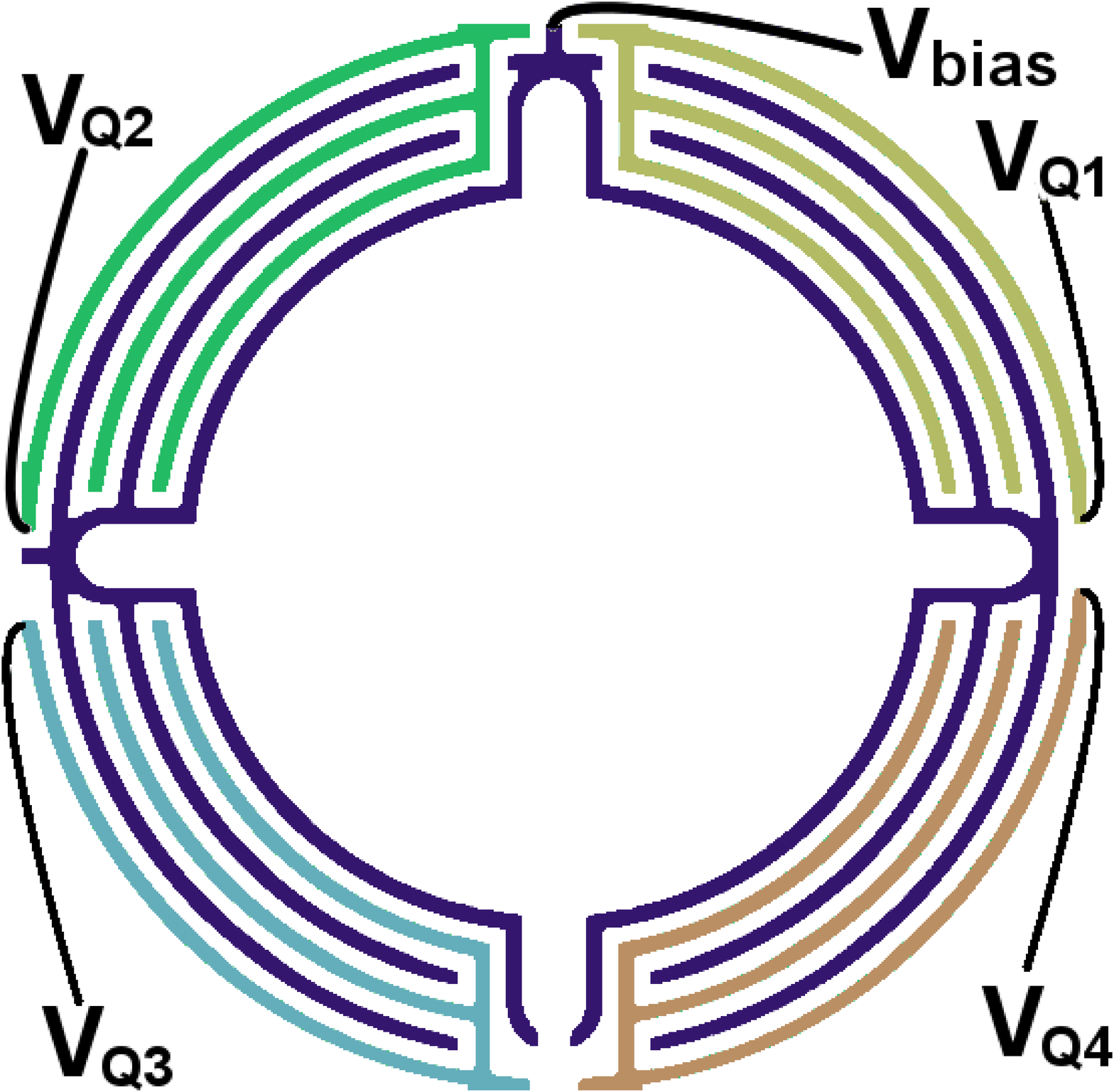}
\centering\includegraphics[width=0.47\linewidth]{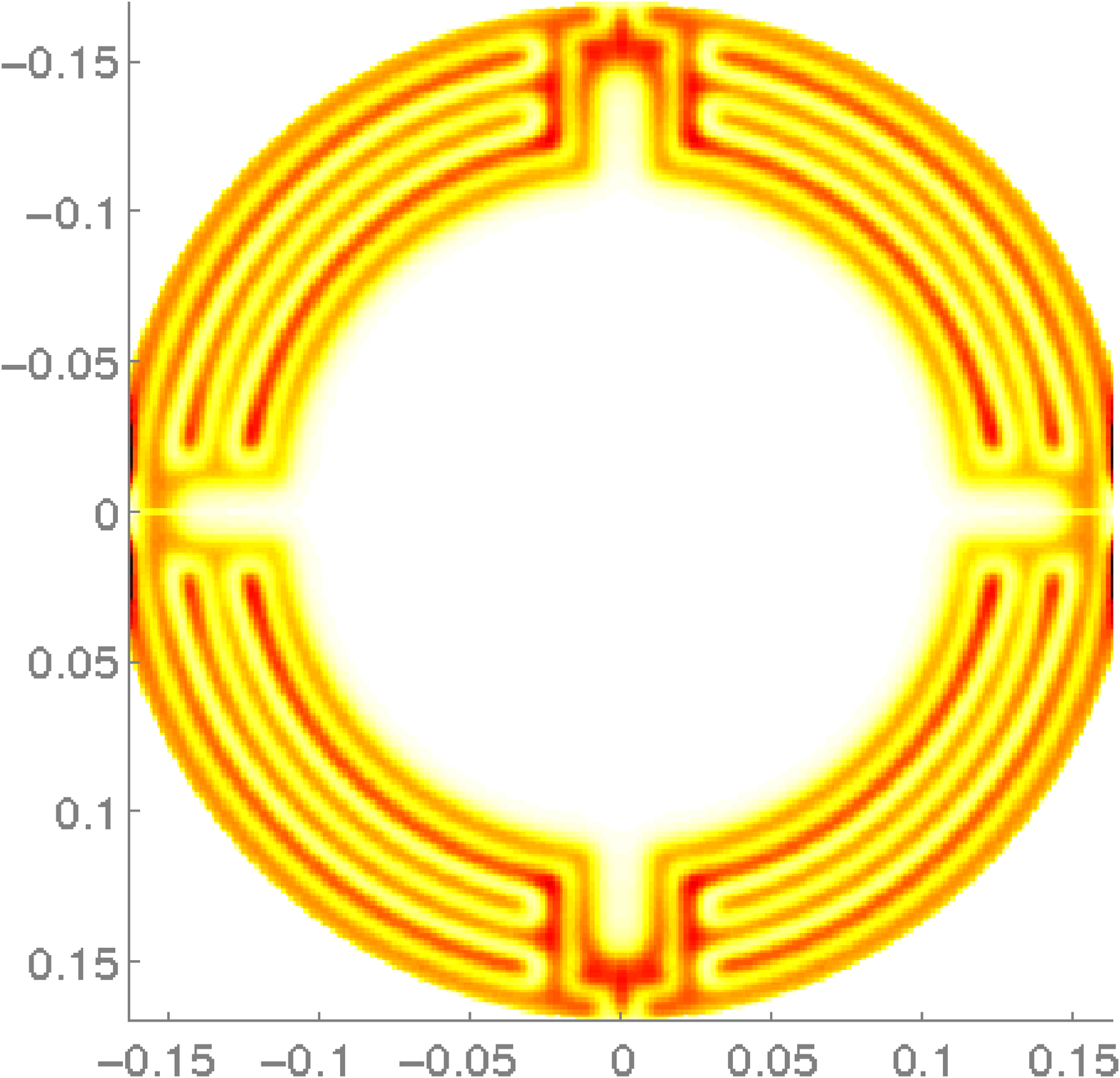}
\caption{The ESD comb pattern printed on the reaction mass (left) and the force distribution on the test mass (right) with the same voltage on all quadrants}\label{fig:ESDMask}
\end{figure}

The main purpose of the electrostatic drive (ESD) is to provide longitudinal actuation on the test masses for lock acquisition \cite{LockAq2006} and holding the arm cavities on resonance.  It creates a force between the test masses and their counterpart reaction masses, through the interaction of the fused silica test masses with the electric fields generated by a comb of gold conductors that are deposited on the reaction mass.  The physical locations of these components are depicted in Figure~\ref{fig:ESDlocation}.  Detail of the gold comb is shown in Figure~\ref{fig:ESDMask} along with the force density on the test mass.  

The force applied to the test mass $F_{ESD}$ is dominated by the dipole attraction of the test mass dielectric to the electric field between the electrodes of the gold comb.  Some portion $b_m$ of this force that couples to the acoustic mode as;
\begin{equation}
F_{\rm app,m} = b_{m} F_{\rm ESD,Q} = b_m \alpha_{\rm Q} \times \frac{1}{2}(V_{\rm bias} - V_{\rm Q})^2\label{eq:QuadForce}
\end{equation}
Here $\alpha_{\rm Q}$ is the force coefficient for a single quadrant, while $V_{\rm bias}$ and $V_{\rm Q}$ are the voltages of the ESD electrodes defined in Figure~\ref{fig:ESDMask}.  
The overlap $b_{\rm m}$ between the ESD force distribution $\vec{f}_{\rm ESD,Q}$ and the displacement $\vec{u}_{\rm m}$ of the surface for a particular acoustic mode $m$ can be approximated as a surface integral derived by Miller \cite{Electrostatic}:
\begin{equation}
b_{\rm m} \approx \Big \vert \iint \limits_{\mathcal{S}} \vec{f}_{\rm ESD,Q} \cdot (\vec{u}_{\rm m}\cdot  \hat z ) \dif \mathcal{S} \Big \vert \label{eq:bm1}
\end{equation}


If a feedback system is created that senses the mode amplitude and provides a viscous damping force using the ESD, the resulting time constant of the mode $\tau_{esd}$ is given by;
\begin{equation}
\tau_{\rm esd} = \Big(\frac{1}{\tau_{m}} +\frac{K_m}{2 \mu_m }\Big)^{-1}
\end{equation}
Here $K_{m}$ is the gain applied between the velocity measurement and the ESD actuation force on a mode with time constant $\tau_{m}$ and effective mass $\mu_m$.  
Reducing the effective time constant lowers the effective parametric gain. 

\begin{equation}
R_{\rm eff} = R_m \times \frac{\tau_{esd}}{\tau_{m}} \end{equation}

The force required to reduce a parametric gain $R_m$ to an effective parametric gain $R_{eff}$ when the mode amplitude is the thermally excited amplitude was used by Miller \cite{Electrostatic} to predict the forces required from the ESD for damping PI,
\begin{equation} F_{\rm req} = \frac{x_m \mu_m \omega_{m}^2}{b_m} \Big(\frac{R_{\rm m} - R_{\rm eff}}{Q_mR_{\rm eff}}\Big) \label{eq:dampingForce} 
\end{equation}

at the thermally excited amplitude 
$x_{\rm m}=\sqrt{k_{\rm B}T/\mu_{\rm m}\omega_{\rm 0,m}^2}$, where $k_{\rm B}$ is the Boltzmann constant and $T$ temperature. 

\begin{figure}[h]
\centering\includegraphics[width=0.99\linewidth]{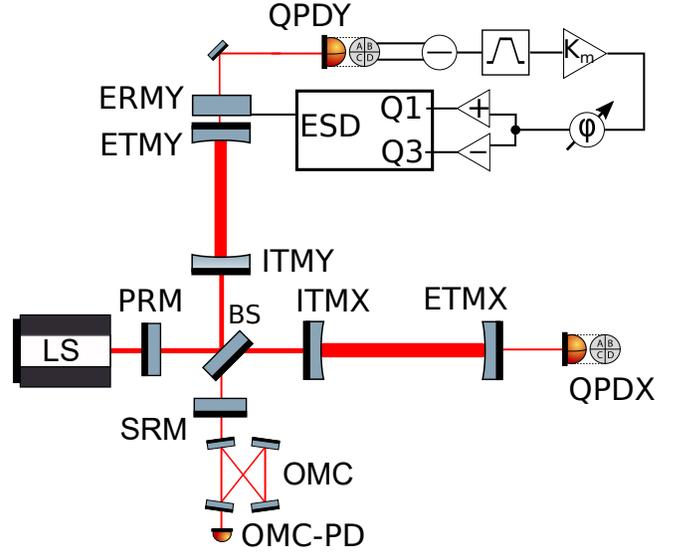}
\caption{A simplified schematic of advanced LIGO showing key components for damping PI in ETMY. Components shown include input and end test masses (ITM/ETM), beam-splitter (BS), power and signal recycling mirrors (PRM/SRM), the laser source (LS), quadrant photo-detectors, the output mode cleaner (OMC), the OMC transmission photo-detector (OMC-PD).  While 4 reaction masses exist, only the Y end reaction mass is shown (ERMY) with key components of the damping loop.  These components generate a differential signal from the vertical orientation of QPDY, filter the signal with a 10\,Hz wide band pass filter centered on the \FDModeFreq\ mode, apply gain $K_{\rm m}$ and phase $\phi$  set in the digital control system and then differentially drive of the upper right $Q1$ and lower left $Q3$ quadrants of the ESD. }\label{fig:LIGOschematic}
\end{figure}

\textit{Feedback Loop} 
Figure~\ref{fig:LIGOschematic} shows the damping feedback loop implemented on the end test mass of the Y-arm (ETMY).  The error signal used for mode damping is constructed from a quadrant photodiode (QPD) that receives light transmitted by ETMY. By suitably combining QPD elements, we measure the beat signal between the cavity $TEM_{\rm 00}$ mode and the $TEM_{\rm 03}$ mode that is being excited by the \FDModeFreq\ ETMY acoustic mode. This signal is band-pass filtered at \FDModeFreq, then phase shifted to produce a control signal that is 90 degrees out of phase with the mode amplitude (velocity damping). The damping force is applied, with adjustable gain, to two quadrants of the ETMY electro-static actuator.   

\begin{figure*}
\centering\includegraphics[width=0.96\linewidth]{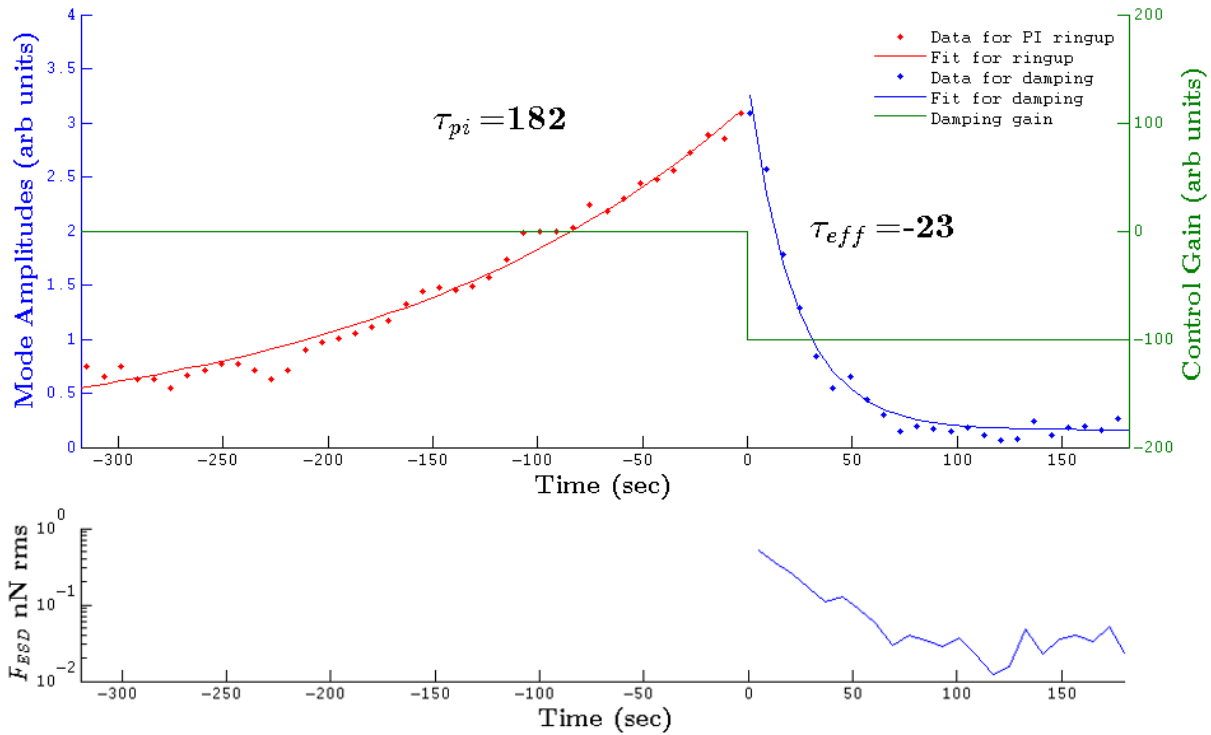}
\caption{Damping of parametric instability.  Upper panel, the \FDModeFreq\ ETMY mode is unstable ringing up with a time constant of \FDRingupTauwU\ and estimated parametric gain of $R_{\rm m}=\FDR$.  Then at 0\,sec control gain is applied resulting in an exponential decay with a time constant of \FDRingdownTauwU\ and effective parametric gain $R_{\rm eff,m}=\FDReff$.  Lower panel, the control force over the same period.  }\label{fig:damping3}
\end{figure*}

\textit{Results} 
PI stabilization via active damping was demonstrated by first causing the ETMY \FDModeFreq\ to become parametrically unstable; this was done by turning off the ring heater tuning, so that the $TEM_{\rm 03}$ mode optical gain curve better overlapped this acoustic mode, as shown in Figure~\ref{fig:modeCartoon}.
When the mode became significantly elevated in the QPD signal, the damping loop was closed with a control gain to achieve a clear damping of the mode amplitude and a control phase optimised to $\pm 15$ degrees of viscous damping.  The mode amplitude was monitored using the photodetector at the main output of the interferometer (labelled OMC-PD in Figure~\ref{fig:LIGOschematic}), as it was found to provide a higher signal-to-noise ratio than the QPD. 

The results are shown in Figure~\ref{fig:damping3}, which plots the mode amplitude during the unstable ring-up phase, followed by the ring-down when the damping loop is engaged. From the ring-up phase, we estimate the parametric gain to be \FDRwU\ from Equation~\ref{eq:piEffQ}.  With the damping applied, 
\begin{equation}
R_{\rm eff} = \frac{R_m \tau_{\rm eff}}{\tau_m+R_m \tau_{\rm eff}} \label{eq:Rcalcs}
\end{equation}
the effective parametric gain is reduced to a stable value of $R_{\rm eff}=$\FDReffwU.  The uncertainty is primarily due to the uncertainty in the estimate of $\tau_m$ which was obtained by the method described in \cite{EvansPIobs}.

At the onset of active damping (time t = 0 in Figure~\ref{fig:damping3}), the feedback control signal produces an estimated force of $F_{esd} =$ \FDStartDampingForce\ (at \FDModeFreq). As the mode amplitude decreased the control force dropped to a steady state value of \FDDampedStateAverageRMS. Over a 20 minute period in this damped state, the peak control force was \FDDampedStatePeak.

\begin{table}
\caption {List of parameters for analysis with values and descriptions} \label{tab:parameters}
\begin{tabular}{ p{1.1cm}p{1.6cm}p{5cm}  }
 \hline
 Symbol  & Value & Description\\
 \hline
 $Q_m$   &  $12\times 10^6$        & Q factor of \FDModeFreq\ mode \\
 P     &  100\,kW  & Power contained in arm cavity   \\
 $\omega_m/2\pi$ & \FDModeFreq\ & Frequency of unstable mode\\
 M  &40kg & mass of test mass\\
  $b_m$  &0.17 & effective mass scaled ESD overlap factor for \FDModeFreq\ mode\\
  $\lambda_0$&   1064\,nm  & laser wavelength\\
 $\alpha_Q$ & $4.8\times 10^{-11} \newline N/V^2$  & ESD quadrant force coefficient   \\
 L & 4km  & Arm cavity length\\
 $V_{\rm bias}$ & 400V  & Bias voltage on ESD\\ 
 $V_{\rm Q}$ & [-20,20]V  & ESD control voltage range\\ 
 \hline 
\end{tabular}
\end{table}

\textit{Discussion} 
The force required to damp the \FDModeFreq\ mode when advanced LIGO reaches design power can be determined from the ESD force used to achieve the observed parametric gain suppression presented here, combined with the expected parametric gain when operated at high power. 

\begin{equation}
\frac{F_{\rm req}}{F_{\rm esd}} = \frac{R_{\rm eff}}{R_{\rm req}}\frac{R_{\rm max} - R_{\rm req}}{R_m - R_{\rm eff}}\label{eq:FracF}
\end{equation}

The maximum parametric gain of the \FDModeFreq\ mode (where $\Delta \omega = 0$) at the power level of these experiments is estimated $\approx 7$ given an estimated de-tuning of $\Delta \omega \approx 50\,Hz$ with zero ring heater power.
At full design power the maximum gain will be $R_{\rm max}\approx 56$.  To obtain a quantitative result, we set a requirement for damping such that the effective parametric gain of unstable acoustic modes after damping be $R_{\rm req} = 0.1$.

Using Equation~\ref{eq:FracF}, the measurements of $R_m$ and $R_{\rm eff}$, the maximum force required to maintain the damped state at high power is $F_{\rm ESD}=$ \FDHighPowerDSAverageRMS.  
Prior to this investigation Miller predicted ~\cite{Electrostatic} that a control force of approximately $10\,\rm nN\, rms$ would be required to maintain this mode at the thermally excited level.    

The PI control system must cope with elevated mode amplitudes as the PI mode may build up before PI control can be engaged.  There is therefore a requirement for some control range or safety factor such that the control system will not saturate if the mode amplitude is a multiple of the safety factor times the damped state amplitude.  
The average ESD drive voltage $V_{Q}$ over the duration the mode was in the damped state was \FDAverageRMSDSVoltage, however during this time it peaked at $\pm$ \FDPeakDSVoltage\ out of a $\pm20\,\rm V$ control range, leading to a safety factor of more than \FDSafetyFactor.  At high power the safety factor will be reduced by the required force ratio of Equation~\ref{eq:FracF} resulting in an expected safety factor of \FDSafetyFactorHP.

As the laser power is increased, other modes are likely to become unstable.  The parametric gain of these modes should be less than the gain of mode group E provided the optical beat note frequency used in these experiments is maintained.  However these modes may also have lower spatial overlap $b_m$ with the ESD.   Miller's simulation \cite{Electrostatic} show some modes in the 30-90\,kHz range will require up to 30 times the control force $F_{ESD}$ required to damp the group E modes.  Even in this situation the PI safety factor is approximately \FDSafetyFactorOM.  

\textit{Conclusion} We have shown for the first time electrostatic control of parametric instability.  An unstable acoustic mode at \FDModeFreq\ with a parametric gain of \FDRwU\ was successfully damped to a gain of \FDReffwU, using electrostatic control forces.   The damping force required to keep the mode in the damped state was \FDDampedStateAverageRMS.  
The prediction through FEM simulation was that the ESD would need to apply approximately six times this control force to maintain the mode amplitude at the thermally excited level.  At high power it is estimated that damping the \FDModeGroupEFreq\ mode group to an effective parametric gain of 0.1 will result in a safety factor  $\approx$ \FDSafetyFactorHP.  It is predicted that unstable modes that are most problematic to damp will still have a safety factor of \FDSafetyFactorOM.  

\textit{Acknowledgments} The authors would like to acknowledge the entire LIGO Scientific Collaboration for the wide ranging expertise that has contributed to these investigations.  
LIGO was constructed by the California Institute of Technology and Massachusetts Institute of Technology with funding from the National Science Foundation, and operates under Cooperative Agreement No. PHY-0757058.  Advanced LIGO was built under Grant No. PHY-0823459.  This paper has LIGO Document Number LIGO-P1600090.
The corresponding author was supported by the Australian Research Council and the LSC fellows program. 

\bibliography{mybib}

\end{document}